\documentclass[preprint,aps,12pt,showpacs,nofootinbib,tightenlines]{revtex4}
\usepackage{amsmath}
\usepackage{amssymb}
\usepackage{epsfig}
\usepackage{graphicx}
\usepackage{subfigure}
\newcommand{\met}{\not\!\!\!E_{T}}
\begin{document}
\def\pslash{\rlap{\hspace{0.02cm}/}{p}}
\def\eslash{\rlap{\hspace{0.02cm}/}{e}}
\title {Single vector-like top partner production in the Left-Right Twin Higgs model at TeV energy $e\gamma$
colliders}
\author{%
GUO Zhan-Ying$^{1}$%
\quad YANG Guang$^{2}$ \quad YANG Bing-Fang
$^{2,3;1)}$\email{yangbingfang@gmail.com} }
\address{%
$^1$Department of Physics, Jiaozuo Normal College, Jiaozuo 454001,
China\\$^2$ Basic Teaching Department, Jiaozuo University, Jiaozuo
454000, China\\$^3$College of Physics and Information Engineering,
Henan Normal University, Xinxiang, 453007, China
   \vspace*{1.5cm}  }

\begin{abstract}
The left-right twin Higgs model contains a new vector-like heavy top
quark, which mixes with the SM-like top quark. In this work, we
studied the single vector-like top partner production via process
$e^{-}\gamma \rightarrow \nu_{e}\bar{T}b$ at the International
Linear Collider. We calculated the production cross section at tree
level and displayed the relevant differential distributions. The
result shows that there will be 125 events produced each year with
$\sqrt{s}$=$2$TeV and the integrated luminosity $\mathcal
L_{int}\simeq 500fb^{-1}$, and the b-quark tagging and the relevant
missing energy $\met$ cut will be helpful to detect this new effect.
\end{abstract}

 \maketitle
\section{ Introduction}
\noindent The top quark was first observed at Ferminlab Tevatron in
1995 \cite{top} and is by far the heaviest elementary fermion. Due
to the large mass, the top quark decays rapidly before forming any
hadronic bound state. Furthermore, the top quark has many properties
different from other quarks, so it occupies a special position in
the standard model(SM) and is often speculated to be sensitive to
the new physics. For these reasons, probing the properties of the
top quark is always one of the forefront topics at the various high
energy collider.

The twin Higgs theories use a discrete symmetry in combination with
an approximate global symmetry to stabilize the Higgs mass. This
mechanism can be implemented in left-right models with the discrete
left-right symmetry \cite{lrhiggs}. In the left-right twin
Higgs(LRTH) model, a vector-like top quark is introduced in order to
give the top quark a mass of the order of electroweak scale. There
is mixing between the SM-like top quark and the heavy top quark so
that the top quark couplings can be modified. At the LHC, a single
vector-like top quark can be produced dominantly via s-channel or
t-channel $W$ or $W_{H}$ exchange, while the production of the
vector-like top quark pair can be produced dominantly from gluon
exchange. The productions and decays of the vector-like top quark at
the LHC have been described in detail in Refs.\cite{heavy top}. If
this effect can be detected, it will be the most compelling evidence
of the new physics.

Duo to the complicated QCD background, the measurement precision of
the LHC is limited. By contrast, the background of the International
Linear Collider(ILC) is very clean so that it will allow unique
opportunities to study the properties and interactions of the SM top
quark and vector-like top quark. Besides the $e^{+}e^{-}$ collider
mode, the $\gamma\gamma$ or $e\gamma$ collider mode can be realized
by the backward Compton scattering at the ILC\cite{ILC}. The search
for deviations from the SM couplings in single top quark production
has become one of the main focus in the on-going and forthcoming
collider experiments\cite{single top}. In this paper, we study the
process $e^{-}\gamma \rightarrow \nu_{e}\bar{T}b$, the results will
be helpful to test the SM and the LRTH model.

This paper is organized as follows. In Sec.II we give a brief review
of the LRTH model. In Sec.III we calculate the production cross
section of the process $e^{-}\gamma \rightarrow \nu_{e}\bar{T}b$ and
the differential distributions of several observables at the ILC.
Finally, we give our conclusions and some comments in Sec.IV.

\section{ A brief review of the LRTH model}
 \noindent
The LRTH model was first proposed in Ref.\cite{LRTH}and some
phenomenological analye and the Feynman rules have been studied in
Ref.\cite{ph-LRTH}. In this section we will briefly review the
essential features of the LRTH model related to our work.

In the LRTH model, the global symmetry is $U(4)\times U(4)$, with a
diagonal subgroup $SU(2)_{L}\times SU(2)_{R}\times U(1)_{B-L}$
gauged. Two Higgs fields, $H$ and $\hat{H}$, are introduced and each
transforms as $(\textbf{4},\textbf{1})$ and
$(\textbf{1},\textbf{4})$ respectively under the global symmetry.
They are written as
\begin{eqnarray}
H=\left( \begin{array}{c} H_{L}\\
H_{R} \\
\end{array}  \right)\,,~~~~~~~~~~~~~~\hat{H}=\left( \begin{array}{c} \hat{H}_{L}\\
\hat{H}_{R} \\
\end{array}  \right)\,,
\end{eqnarray}
where $H_{L,R}$ and $\hat{H}_{L,R}$ are two component objects which
are charged under the $SU(2)_{L}\times SU(2)_{R}\times U(1)_{B-L}$
as
\begin{equation}
H_{L}{\rm \ and\ }\hat{H}_{L}: (\textbf{2}, \textbf{1},
1),~~~~~~~~H_{R}{\rm \ and\ } \hat{H}_{R}: (\textbf{1}, \textbf{2},
1).
\end{equation}
After Higgses develop vacuum expectation values (vevs) as $\langle
H\rangle=(0,0,0,f)$ and $\langle \hat{H}\rangle=(0,0,0,\hat{f})$,
which break the $U(4)\times U(4)$ global symmetry as well as the
gauge symmetry ${\rm SU}(2)_R \times {\rm U}(1)_{B-L}$ down to the
SM ${\rm U}(1)_Y$.

After the electroweak breaking of $SU(2)_{L}\times U(1)_{Y}$, three
Goldstone bosons are eaten by the massive gauge bosons $W^{\pm}$ and
$Z$ in the SM, their masses can be given by
\begin{eqnarray}
M_{W}^{2}&=& \frac{1}{2}g_{2}^{2}f^{2}\sin^{2}x, \\
M_{Z}^{2}&=&\frac{g_{2}^{2}+g_{Y}^{2}}{g_{2}^{2}}\frac{2M_{W}^{2}M_{W_{H}}^{2}}{M_{W}^{2}+M_{W_{H}}^{2}+\sqrt{(M_{W_{H}}^{2}-M_{W}^{2})^{2}+4\frac{g_{1}^{4}}{(g^{2}+g_{1}^{2})^{2}}M_{W_{H}}^{2}M_{W}^{2}}}
\end{eqnarray}
where $x=v/(\sqrt{2}f)$ and $v=246$GeV is the electroweak scale,
$g_{Y}$ is the SM hypercharge coupling, $M_{W_{H}}$ is the
$W_H^{\pm}$ mass, the values of $f$ and $\hat{f}$ will be bounded
from below by electroweak precision measurements.

In order to give the top quark mass of the order of the electroweak
scale, a pair of vector-like quarks $Q_{L}$ and $Q_{R}$ is
introduced. The mass eigenstates, which contain one of the SM-like
top quark $t$ and a heavy top partner $T$, are mixtures of the gauge
eigenstates. Their masses are given by
\begin{eqnarray}
m_{t}^{2} = \frac{1}{2}(M^{2}+y^{2}f^{2}-N_{t}),\hspace{0.5cm}
M_{T}^{2} = \frac{1}{2}(M^{2}+y^{2}f^{2}+N_{t}),
\end{eqnarray}
where $N_{t} = \sqrt{(y^{2}f^{2}+M^{2})^{2}-y^{4}f^{4}\sin^{2}2x}$.

At the leading order, the mixing angles for the left-handed and
right-handed fermions are
\begin{eqnarray}
s_{L}&\simeq& \frac{M}{m_{T}}\sin x,\\
s_{R}&\simeq& \frac{M}{m_{T}}(1+\sin^{2}x),
\end{eqnarray}
where $M$ is the mass parameter essential to the mixing between the
top quark $t$ and its partner $T$.

\section{Single vector-like top production via process $e^{-}\gamma \rightarrow \nu_{e}\bar{T}b$}
\begin{figure}[htbp]
\begin{center}
\includegraphics[width=11cm]{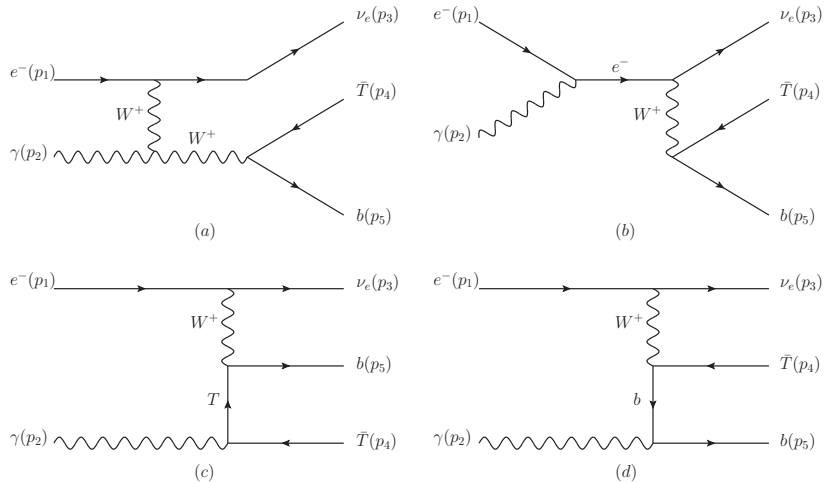}\vspace{-0.5cm}
\caption{Feynman diagrams of the process $e^{-}\gamma \rightarrow
\nu_{e}\bar{T}b$ in the left-right twin Higgs model.}
\end{center}
\end{figure}

\noindent  At a linear collider the single vector-like top quarks
can be produced from the following two processes:
\begin{eqnarray}
T:e^{+}\gamma \rightarrow
\bar{\nu}_{e}T\bar{b},~~~~~~~~~\bar{T}:e^{-}\gamma \rightarrow
\nu_{e}\bar{T}b
\end{eqnarray}
where the photon comes from the original incoming electron and
positron, respectively. The relevant Feynman diagrams of the process
$e^{-}\gamma \rightarrow \nu_{e}\bar{T}b$ in the LRTH model are
shown in Fig.1.

The invariant production amplitudes of the process $e^{-}\gamma
\rightarrow \nu_{e}\bar{T}b$ can be written as:
\begin{equation}
\mathcal{M}= \mathcal{M}_{a}+
\mathcal{M}_{b}+\mathcal{M}_{c}+\mathcal{M}_{d}
 \end{equation}
 with
\begin{eqnarray}
 M_{a}&=&\varepsilon_{\rho}(p_{2})\bar{u}(p_{3})V_{W\nu e}^{\mu}u(p_{1})\bar{u}(p_{5})V_{WTb}^{\alpha}v(p_{4})V_{WW\gamma}^{\nu\rho\beta}\nonumber\\
 &&\frac{-ig_{\mu\nu}}{(p_{1}-p_{3})^{2}-m_{W}^{2}}\frac{-ig_{\alpha\beta}}{(p_{4}+p_{5})^{2}-m_{W}^{2}}\\
 M_{b}&=&\varepsilon_{\rho}(p_{2})\bar{u}(p_{3})V_{W\nu e}^{\mu}\frac{i}{(\pslash_{1}+\pslash_{2})-m_{e}}V_{\gamma ee}^{\rho}u(p_{1})\bar{u}(p_{5})V_{WTb}^{\nu}v(p_{4})\nonumber\\
 &&\frac{-ig_{\mu\nu}}{(p_{4}+p_{5})^{2}-m_{W}^{2}}\\
M_{c}&=&\varepsilon_{\rho}(p_{2})\bar{u}(p_{3})V_{W\nu e}^{\mu}u(p_{1})\bar{u}(p_{5})V_{WTb}^{\nu}\frac{i}{(\pslash_{2}-\pslash_{4})-m_{T}}V_{\gamma TT}^{\rho}v(p_{4})\nonumber\\
 &&\frac{-ig_{\mu\nu}}{(p_{1}-p_{3})^{2}-m_{W}^{2}}\\
M_{d}&=&\varepsilon_{\rho}(p_{2})\bar{u}(p_{3})V_{W\nu e}^{\mu}u(p_{1})\bar{u}(p_{5})V_{\gamma bb}^{\rho}\frac{i}{(\pslash_{5}-\pslash_{2})-m_{b}}V_{WTb}^{\nu}v(p_{4})\nonumber\\
 &&\frac{-ig_{\mu\nu}}{(p_{1}-p_{3})^{2}-m_{W}^{2}}
\end{eqnarray}
where $V$ denotes the three-point vertices of the particles, the
relevant Feynman rules can be found in Ref.\cite{ph-LRTH}.

With the above amplitudes, we can directly obtain the production
cross section $\hat{\sigma}(\hat{s})$ for the subprocess
$e^{-}\gamma \rightarrow \nu_{e}\bar{T}b$, which can be obtained by
folding $\sigma(\hat{s})$ with the photon distribution
function\cite{function}:
\begin{equation}
\sigma(tot)=\int^{x_{max}}_{(m_{T}+m_{b})^{2}/s}dx\sigma(\hat{s})f_{\gamma}(x),
 \end{equation}
where
\begin{equation}
f_{\gamma}(x)=\frac{1}{D(\xi)}[1-x+\frac{1}{1-x}-\frac{4x}{\xi(1-x)}+\frac{4x^{2}}{\xi^{2}(1-x)^{2}}],
\end{equation}
with
\begin{equation}
D(\xi)=(1-\frac{4}{\xi}-\frac{8}{\xi^{2}})\ln(1+\xi)+\frac{1}{2}+\frac{8}{\xi}-\frac{1}{2(1+\xi)^{2}}.
\end{equation}
where $\xi=\frac{4E_0\omega_{0}}{m^{2}_{e}}$, $E_0$ and $\omega_0$
are the incident electron and laser light energies, and
$x=\omega/E_0$. $f_{\gamma}$ vanishes for
$x>x_{max}=\omega_{max}/E_{e}=\xi/(1+\xi)$. We require
$\omega_{0}x_{max}\leq m_{e}^{2}/E_{e}$, which implies that $\xi\leq
2+2\sqrt{2}\simeq4.8$. We choose $\xi=4.8$, then obtain
\begin{equation}
x_{max}\approx0.83,~~~~~~~~~~D(\xi_{max})\approx1.8.
 \end{equation}
For simplicity, the possible polarization for the electron and
photon beams have been ignored, and the number of the backscattered
photons produced per electron is assumed to be one.

We take the SM parameters used in our calculations as\cite{para}
\begin{eqnarray}
\nonumber &&G_{F}=1.16637\times 10^{-5}\textmd{GeV}^{-2},
S^{2}_{W}=0.231,
\alpha_{e}=1/128,\\&&M_{Z}=91.2\textmd{GeV},m_{t}=174.3\textmd{GeV},\Gamma
_{Z}=2.436\textmd{GeV}.
\end{eqnarray}

The relevant LRTH parameters in our calculation are the scale $f$,
the mass parameter $M$ and the heavy top quark mass $m_{T}$.
Recently, the ATLAS Collaboration presented a search that
vector-like top quark with mass lower than 656 GeV is excluded at
95\% confidence level\cite{ATLAS}. Earlier, the CMS Collaboration
presented a search that vector-like top quark mass below 557GeV is
excluded at 95\% confidence level\cite{CMS}. Considering these
constraints to the relevant LRTH parameters, we take $M=150$GeV and
make the $m_{T}$ and $f$ satisfy Eq. (5) at all times.

\begin{figure}[htbp]
\begin{center}
\includegraphics[width=7cm]{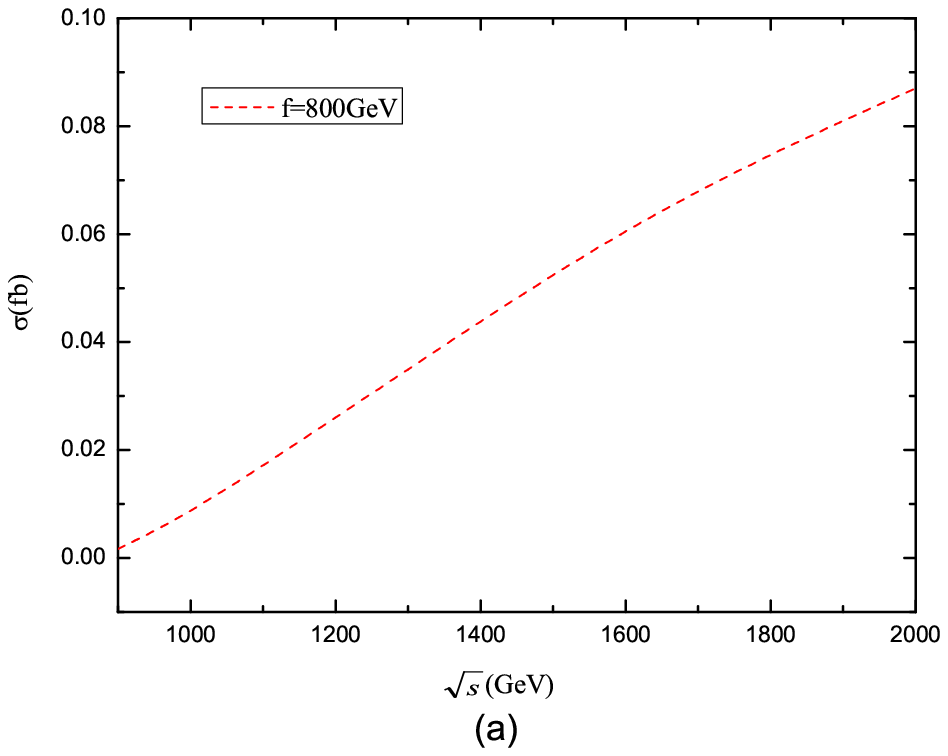}\hspace{-0.8cm}
\includegraphics[width=7cm]{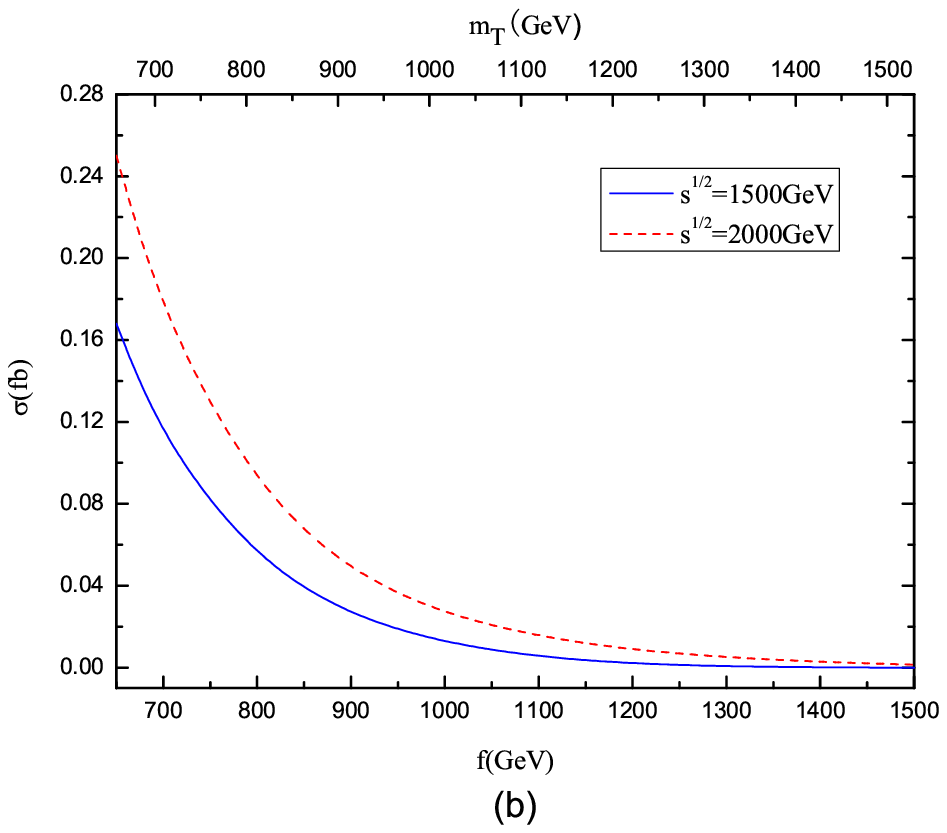}\vspace{-0.8cm}
\caption{The production cross section $\sigma$ as functions of the
center-of-mass energy $\sqrt{s}$ (a) and the scale $f$(b).}
\end{center}
\end{figure}

In Fig.2(a), we discuss the dependance of the production cross
section $\sigma$ on the center-of-mass energy $\sqrt{s}$ for
$f=800$GeV. We can see that the cross section $\sigma$ becomes lager
with the $\sqrt{s}$ increasing.

In Fig.2(b), we discuss the dependance of the production cross
section $\sigma$ on the scale $f$ for
$\sqrt{s}=1500\textmd{GeV},2000\textmd{GeV}$, respectively. We can
see the $\sigma$ decreases as the scale $f$ increases, which means
that the vector-like top quark production cross section decouples
with the scale $f$ increasing. The maximum of the production cross
section can reach 0.17fb for $\sqrt{s}=1500$GeV and 0.25fb for
$\sqrt{s}=2000$GeV, respectively. If we take the integrated
luminosity $\mathcal L_{int}\simeq 500fb^{-1}$, there will be 125
events produced each year with $\sqrt{s}$=$2$TeV.

In Fig.3, we display the normalized transverse momentum
distribution, the normalized distribution for the missing energy and
the normalized rapidity of bottom quark of the process $e^{-}\gamma
\rightarrow \nu_{e}\bar{t}b$ in the SM and the process $e^{-}\gamma
\rightarrow \nu_{e}\bar{T}b$ in the LRTH model.

\begin{figure}[htbp]
\begin{center}
\includegraphics[width=4.7cm]{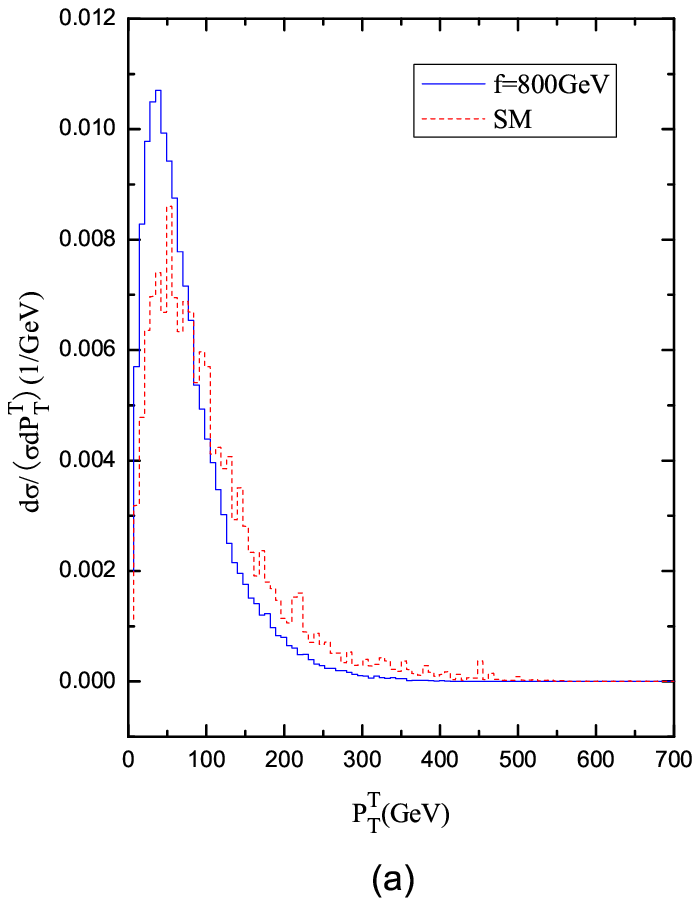}\hspace{-0.3cm}
\includegraphics[width=4.8cm]{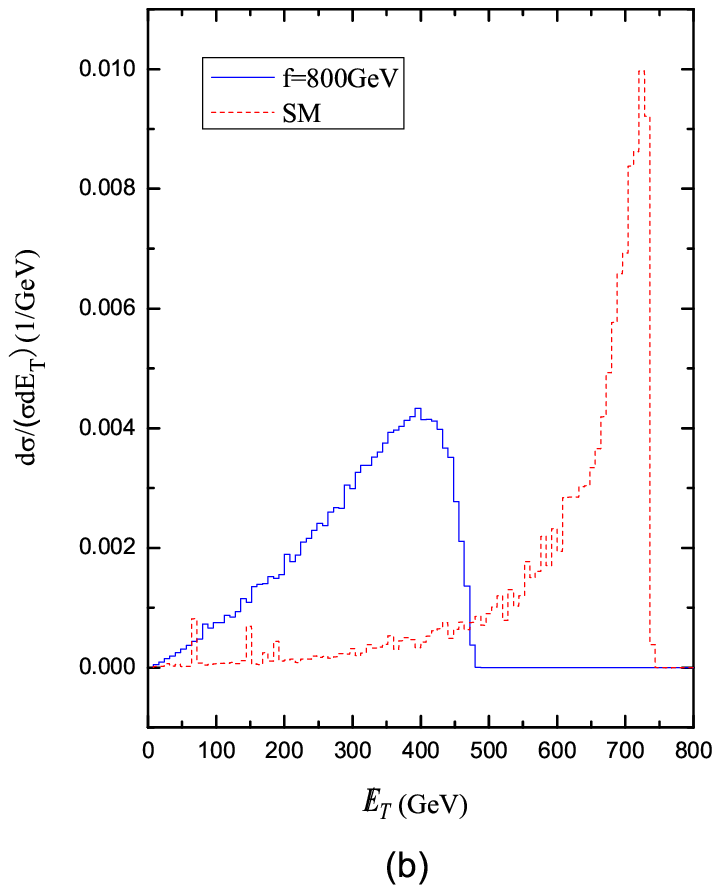}\hspace{-0.3cm}
\includegraphics[width=4.8cm]{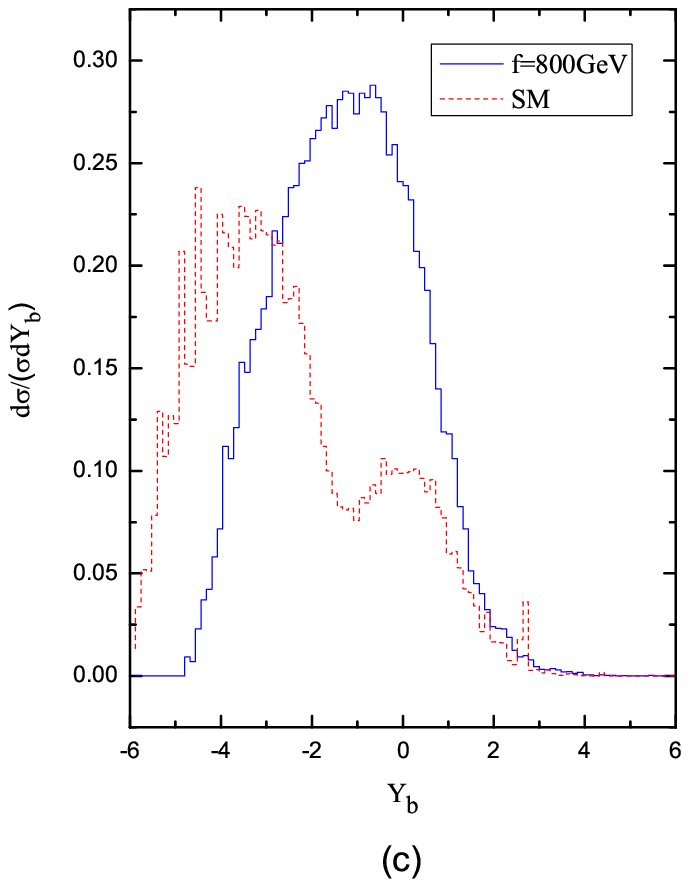}\vspace{-0.6cm}
\caption{Normalized differential distributions for the SM top quark
and the heavy top quark transverse momentum (a), the missing energy
(b) and the rapidity of the bottom quark (c) for
$\sqrt{s}=1500$GeV.}
\end{center}
\end{figure}

In Fig.3(a), we show the normalized transverse momentum distribution
behaviour of the SM top quark and the heavy top quark. As the
neutrino comes from the initial state positron after emitting a $W$
boson, the heavy top quark transverse momentum peaks at $\sim
M_{W}/2$, which is very similar to the transverse momentum
distribution behaviour of the SM top quark.

In Fig.3(b), we show the normalized distribution for the missing
energy $\met$ carried by the final-state neutrino. Compared with the
SM top quark, the peak of the normalized distribution moves to the
low energy region. With the scale $f$ increasing, this peak of the
normalized distribution moves to the left. If we take a relevant
missing energy $\met$ cut, such as $\met < 500$GeV for $f=800$GeV,
the SM background can be suppressed effectively. Considering the
subsequent decay of the heavy top quark, the main signal is 4 b jets
+ one charged lepton (e or $\mu$) + energy $\met$. Because the
additional b jet carries off energy, the peak of the missing energy
$\met$ in the LRTH model is lower than the peak in the SM.

In Fig.3(c), we show the normalized rapidity distribution of bottom
quark. The $WW\gamma$ diagram (Fig.1(a))corresponds to a virtual $W$
boson moving in the positive rapidity region to balance the $\nu$
emitted from the incoming $e^{-}$. This virtual $W$ boson's decay
products, the $b$ and $\bar{t}$ quarks, lead to the small kink in
the right region. By contrast, the same thing happens in the process
$e^{-}\gamma \rightarrow \nu_{e}\bar{T}b$, but the difference is
that the kink is smaller due to the large mass of the heavy top
quark.

\section{Conclusions} \noindent In this paper, we studied the single vector-like top production
process $e^{-}\gamma \rightarrow \nu_{e}\bar{T}b$ in the LRTH model.
The result shows that there will be 125 events produced each year
with $\sqrt{s}$=$2$TeV and the integrated luminosity $\mathcal
L_{int}\simeq 500fb^{-1}$ at small value of the scale $f$ around 650
GeV. Now, by the b-quark tagging efficiency of
$70\%$\cite{b-tagging} and the relevant missing energy $\met$ cut,
this new effects will help to test the LRTH model and probe the new
physics at the ILC.

\end{document}